\documentclass[fleqn,usenatbib]{mnras}

\usepackage{newtxtext,newtxmath}
\usepackage[T1]{fontenc}

\DeclareRobustCommand{\VAN}[3]{#2}
\let\VANthebibliography\thebibliography
\def\thebibliography{\DeclareRobustCommand{\VAN}[3]{##3}\VANthebibliography}

\usepackage{graphicx}	% Including figure files
\usepackage{amsmath}	% Advanced maths commands

\title[Swift Analysis]{A Swift analysis of the Eras tour set list and implications for astrophysics research (Taylor's version)}

\author[S. L. Newman et al.]{
Sophie L. Newman,$^{1}$\thanks{E-mail: sophie.newman@port.ac.uk}
Ana Sainz de Murieta,$^{1}$
\\
$^{1}$Institue of Cosmology \& Gravitation, University of Portsmouth, UK
}

\date{Accepted XXX. Received YYY; in original form ZZZ}

\pubyear{\the\year{}}

\begin{document}
\label{firstpage}
\pagerange{\pageref{firstpage}--\pageref{lastpage}}
\maketitle

% Abstract of the paper
\begin{abstract}
    Popular culture plays a significant role in shaping public interest in science, and Taylor Swift’s discography frequently incorporates astrophysics terminology. This study examines the occurrence of astrophysics-related words in her lyrics and their representation in the Eras Tour set list. By analyzing the frequency of words in Swift's total discography, we identify that astrophysics is promoted the most within her most recent album, \textit{The Tortured Poets Department}, whereas songs from \textit{Midnights} promoted astrophysics the most throughout the Eras tour. We categorize words into various disciplines of astrophysics and find that multimessenger astronomy is promoted the most, both in Swift's total discography and throughout the Eras tour. We perform a Taylor expansion and predict $12 \pm 5$ astrophysical terms in Swift's next album. This analysis offers a unique perspective on the intersection of music and science, revealing how Swift’s artistry may unintentionally promote interest in different fields of astrophysics.
\end{abstract}

% Select between one and six entries from the list of approved keywords.
% Don't make up new ones.

%%%%%%%%%%%%%%%%%%%%%%%%%%%%%%%%%%%%%%%%%%%%%%%%%%

%%%%%%%%%%%%%%%%% BODY OF PAPER %%%%%%%%%%%%%%%%%%

\section{Introduction}

Popular culture has long played a crucial role in shaping public engagement with science. From classic science-fiction literature \citep{herbert,adams,wells} to blockbuster films and chart-topping songs \citep{buchan,asapscience}, artistic expressions often introduce audiences to scientific concepts in ways that formal education or academic discourse may not. Music, in particular, possesses a unique ability to weave complex themes into accessible narratives, using metaphor and emotion to spark curiosity. While previous research has examined the influence of music on social justice issues \citep{socialjustice}, identity formation \citep{identity} and historical memory \citep{history}, its potential to introduce listeners to scientific concepts — especially in fields such as astrophysics — remains a relatively unexplored area of study \citep{subatomic,astrobeats}.

Taylor Swift, one of the most influential and commercially successful artists of the 21st century, frequently incorporates celestial imagery into her songwriting. Words such as "stars," "galaxies," "moonlight," and "cosmos" appear throughout her discography, often serving as metaphors in her many songs. These references, while primarily poetic, provide an intriguing opportunity to examine how popular music might indirectly promote interest in astrophysical concepts. Given Swift’s vast global reach and dedicated fanbase, her lyrical choices may contribute to shaping public perceptions of space and the universe — whether consciously or not.

Beyond her recorded music, Swift’s Eras Tour presents a compelling case study for how these celestial themes are represented in live performance. With each "era" meticulously designed to reflect different stages of her career, the setlist offers insight into which songs and, by extension, which astrophysical references are emphasized. By analyzing the frequency of astrophysics-related words in Swift’s discography and comparing their presence in the Eras Tour, this paper seeks to identify patterns in how cosmic imagery is woven into the highest grossing tour of all time. 

This paper is structured as follows. In Section \ref{sec:ErasTour},  we provide an overview of the Eras Tour and highlight its key moments. Section \ref{sec:data} describes our data sources and outlines the methodology used in our analysis.   In Section \ref{sec:results1}, we identify the songs that contain the most astrophysical words, while Section \ref{sec:eras_tour_promotion} examines the specific areas of astrophysics most prominently featured in the Eras Tour. We present a Taylor Expansion of our results in Section \ref{sec:taylorexpansion} and compare the impact of Taylor Swift on promoting astrophysics compared to other artists in Section \ref{sec:kanye}. Suggestions for an expansion on topics in Swift's lyrics are given in Section \ref{sec:discussion}. Finally, our conclusions are summarised in Section \ref{sec:conclusions}.

\section{The Eras Tour}
\label{sec:ErasTour}

\begin{figure*}
    \centering
    \includegraphics[width=\textwidth]{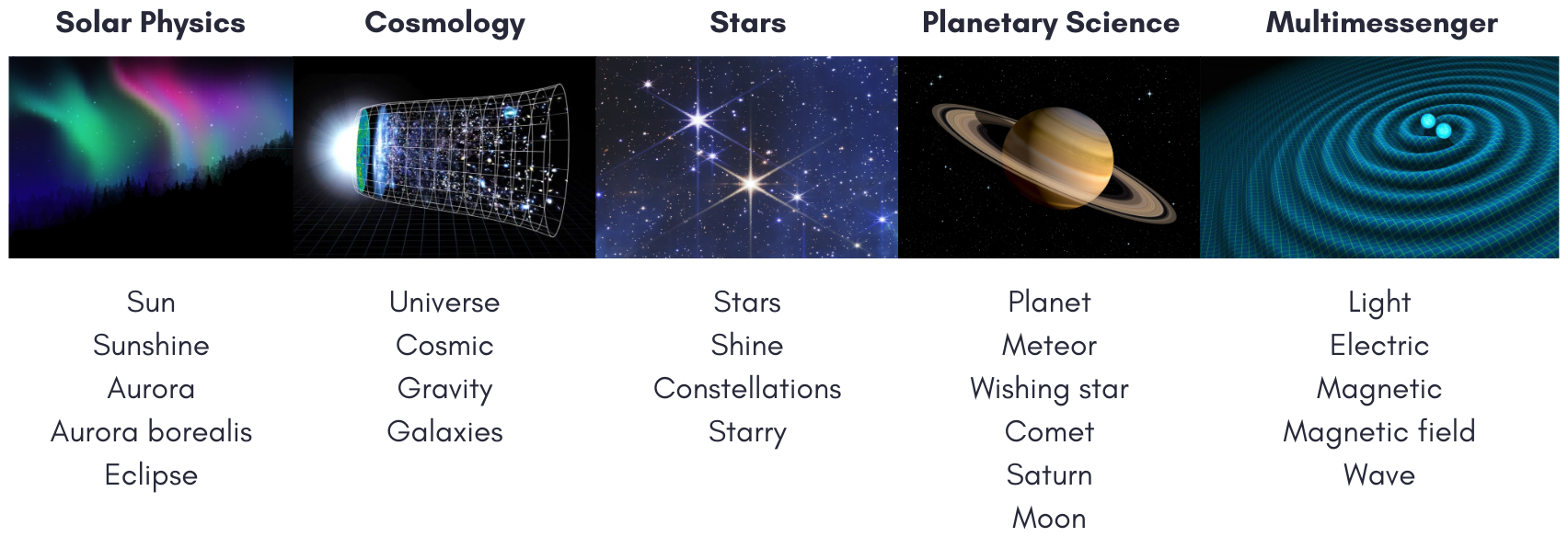}
    \caption{Words from within Swift's discography across different astrophysics disciplines: solar physics, cosmology, stars, planetary science, and multimessenger astronomy.}
    \label{fig:words}
\end{figure*}

Taylor Swift's Eras Tour was her sixth concert tour, kicking off in Glendale, Arizona, on March 17, 2023, and wrapping up in Vancouver, British Columbia, on December 8, 2024 \citep{startend}. With 149 performances in 51 cities across five continents \citep{totalshows}, the tour left a significant cultural and economic impact. It became the highest-grossing tour in history and the first to surpass both $1$ billion and $2$ billion in revenue \citep{gross}. 

Swift designed the tour as a retrospective celebration of her studio albums and their distinct musical "eras" \citep{eras}. Lasting over 3.5 hours, the set list featured more than 40 songs, divided into 10 acts that captured the essence and aesthetic of each album. In May 2024, the show was revamped to include her eleventh studio album, \textit{The Tortured Poets Department} \citep{tortured}.

Iconic moments from the Eras tour include Travis Kelce involved with the performance of `I Can Do It With a Broken Heart' in Wembley \citep{travis}, the blending of seemingly unrelated songs like `Getaway Car,' `August,' and the `The Other Side of the Door' \citep{mashup}, Swift changing the lyrics to `Karma' \citep{karma}, and  $25,000 - 40,000$ fans gathering on a hilltop outside Olympic Stadium in Munich, Germany to watch the show for free \citep{hill}.

\section{Data}
\label{sec:data}

\begin{figure*}
    \centering
    \includegraphics[width=\textwidth]{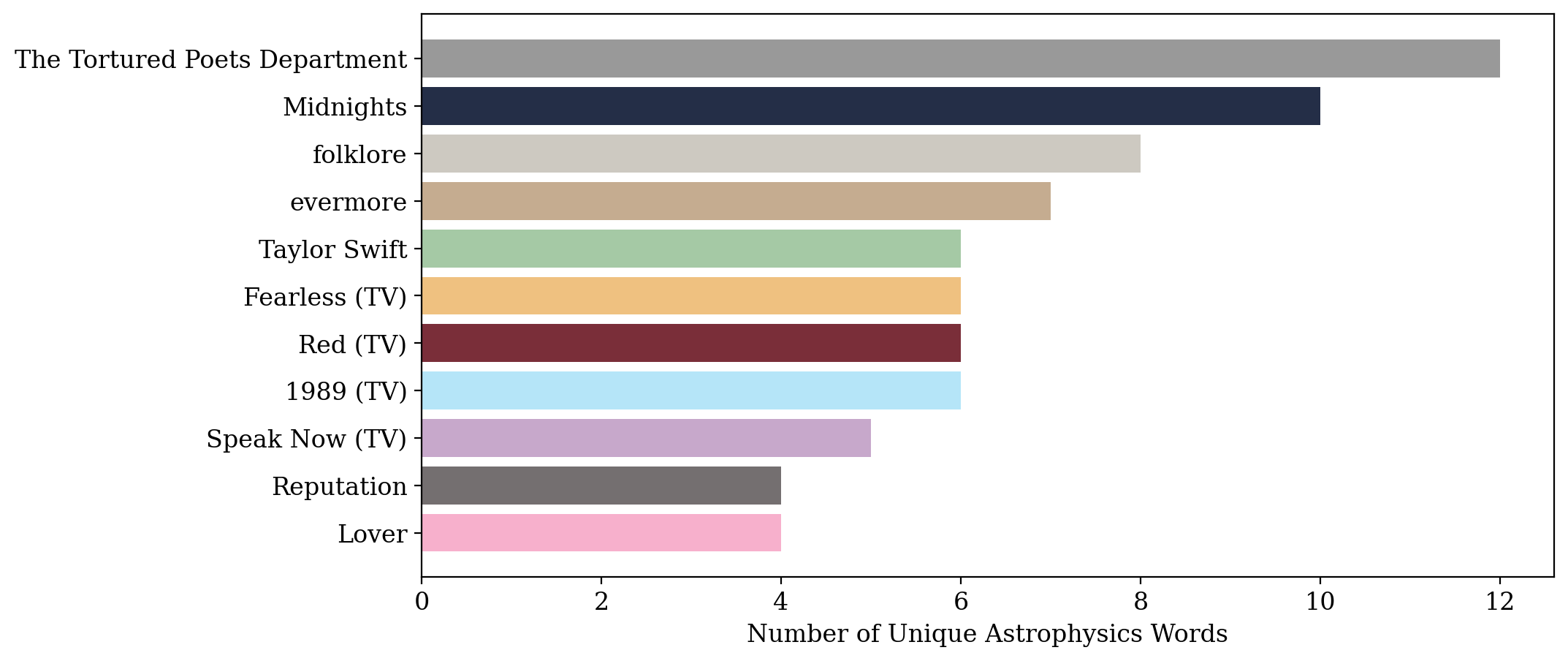}
    \caption{Number of unique astrophysics words per album in Swift's discography}
    \label{fig:albums_vs_words}
\end{figure*}

\begin{figure*}
    \centering
    \includegraphics[width=\textwidth]{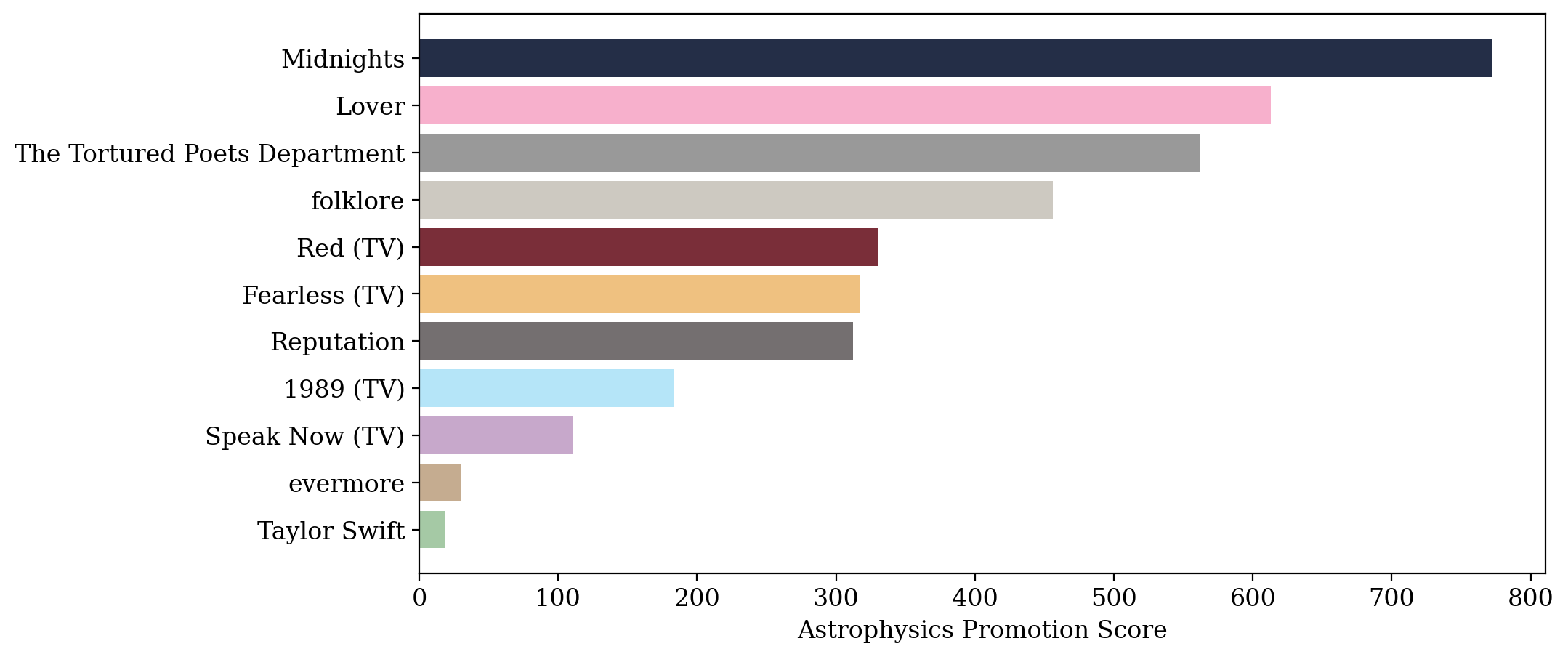}
    \caption{Astrophysics promotion score for each of Swift's albums during the Eras tour, calculating by weighting the number of astrophysics words in each song by the number of times it was played during the Eras tour, then computing the weighted sum for each album.}
    \label{fig:tour_score}
\end{figure*}

We use \texttt{setlist.fm}\footnote{\url{https://www.setlist.fm/stats/taylor-swift-3bd6bc5c.html?tour=6bde5e4e}} which lists how often a song was performed during the Eras tour. Rather than defining set list as a list of songs that an artist \textit{intends} to play, \texttt{setlist.fm} takes that definition further and considers a setlist to be `the list of the songs a band or artist \textit{actually} played during a concert'. If a song is combined with another (e.g. a mashup of `The Archer' and `You're on Your Own, Kid') we count both songs as being played within that set once. 

To search lyrics for astronomy-related words, we employ two different techniques:

\begin{enumerate}
    \item We use a \href{https://shaynak.github.io/taylor-swift/}{lyrics database} made by Shayna Kothari to search for various terms that are important in the astronomy community, such as `Sun' and `gala' for `galaxy', `galaxies', or `galactic'.    
    \item We go through lyrics manually and search for words that we might not have thought about searching for. E.g. through this technique the word `shine' was noticed to occur many times.
\end{enumerate}

To ensure a fair comparison between albums, we count each word only once per song, even if it appears multiple times, as it may be repeated in the chorus. For example, `shine' appears four times in the song `Ours (Taylor's Version)' as part of the chorus. For words that appear in both singular and plural forms (e.g. `light' and `lights'), we make the plural word singular when counting word frequencies. 

In order to determine which field of astrophysics Taylor Swift promotes the most, we characterize each of the words we found in the lyrics into five different categories: solar physics, cosmology, stars, planetary science, and multimessenger \citep[see e.g.][]{sergi,elena} astronomy. Our results are shown in Figure \ref{fig:words}. Our analysis shows that Taylor Swift's lyrics use a larger number of distinct words from the field of planetary science (e.g. planet, Saturn), but this conclusion could change depending on the category assigned to each word. For example, we assigned "eclipse" to the solar physics category, but she could instead be referring to eclipsing binaries, in which case it would belong to the stars category. We conclude that for each of our chosen categories there are $5\pm1$ distinct words found in Taylor Swift's lyrics.

\section{So High School}
\label{sec:results1}

Which albums have the most astrophysics and is the most `High School'? In Figure \ref{fig:albums_vs_words} we show the total number of unique astrophysics terms for each of Taylor Swift's albums. We find the average Swift album has an average of $6.0 \pm 2.3$ unique astrophysics words. We also find that Swift's newest album, The Torturted Poets Department, with 12 unique astrophysics words and therefore $2.6 \sigma$ away from the median, promotes astrophysics research the most to her fans. \textit{Reputation} and \textit{Lover}, on the other hand, promote astrophysics the least out of her whole discography, but do not deviate enough from the median to conclude that these albums do not promote astrophysics research. Interestingly, \textit{folklore} and \textit{evermore}, often called `sister' albums \citep{sister} differ only by one word. We also note that her last 4 original albums (not including re-recordings) contain an increase in astronomical references with respect to her previous releases, which could indicate a more recent interest in the field. This could be explained by recent developments and advancements in the topic over the last 5 years \citep{eventhorizon,jwst,desi}. We note that \textit{folklore}, her eigth studio album, was released in July 2024, amidst the `Covid-19' pandemic. The singer wrote in a note to her fans \citep{folklore} that "in isolation my imagination has run wild and this album is the result". We conclude that it is possible that she became more interested in topic during that period. Moreover, during the first two years of the pandemic, just as Swift did, the field of astronomy experienced a boost in the number of papers published, mainly driven by boosted individual productivity \citep{productivity}. These two factors could have accounted for her increased interest in the field.

\section{...Ready for it? What field of astrophysics did the Eras tour promote?}
\label{sec:eras_tour_promotion}
To calculate which albums promoted astrophysics the most \textit{throughout} the Eras tour, we took the following steps:

\begin{enumerate}
    \item Filter out songs without astrophysical words
    \item Count the number of astrophysical words per song
    \item Compute the astrophysics promotion score for each song as the number of unique astrophysics words within the song multiplied by the number of times the song was played during the Eras tour
    \item Sum these scores for each album
\end{enumerate}

The sum of these scores is denoted as the Astrophysics Promotion Score (APS). The results of these steps are in Figure \ref{fig:tour_score}, where we find the album \textit{Midnights} to have the highest APS out of all albums. This is not surprising as it is the second album with the most unique astrophysics words and the Eras Tour was announced shortly after the release of \textit{Midnights}, meaning there was a large focus on the tour around this album, as it had not been performed live before. We find the album with the lowest APS score to be her debut album, mainly due to the lowest number of songs from this album in the setlist. It appears surprising that \textit{evermore}, which was the fourth album with the highest number of astrophysics-related words, has the second lowest APS score, scoring below 100. However, as other fans have already pointed out, Swift is known to "hate" this album. Even if the singer has denied these claims \citep{buzzfeed}, its low APS score might reflect the singer's personal opinion of the album. Within the Appendix, in Figure \ref{fig:top_20}, we also show the 20 songs that Swift played the most during the Eras tour which contain astrophysical terminology.

\begin{figure*}
    \centering
    \includegraphics[width=\textwidth]{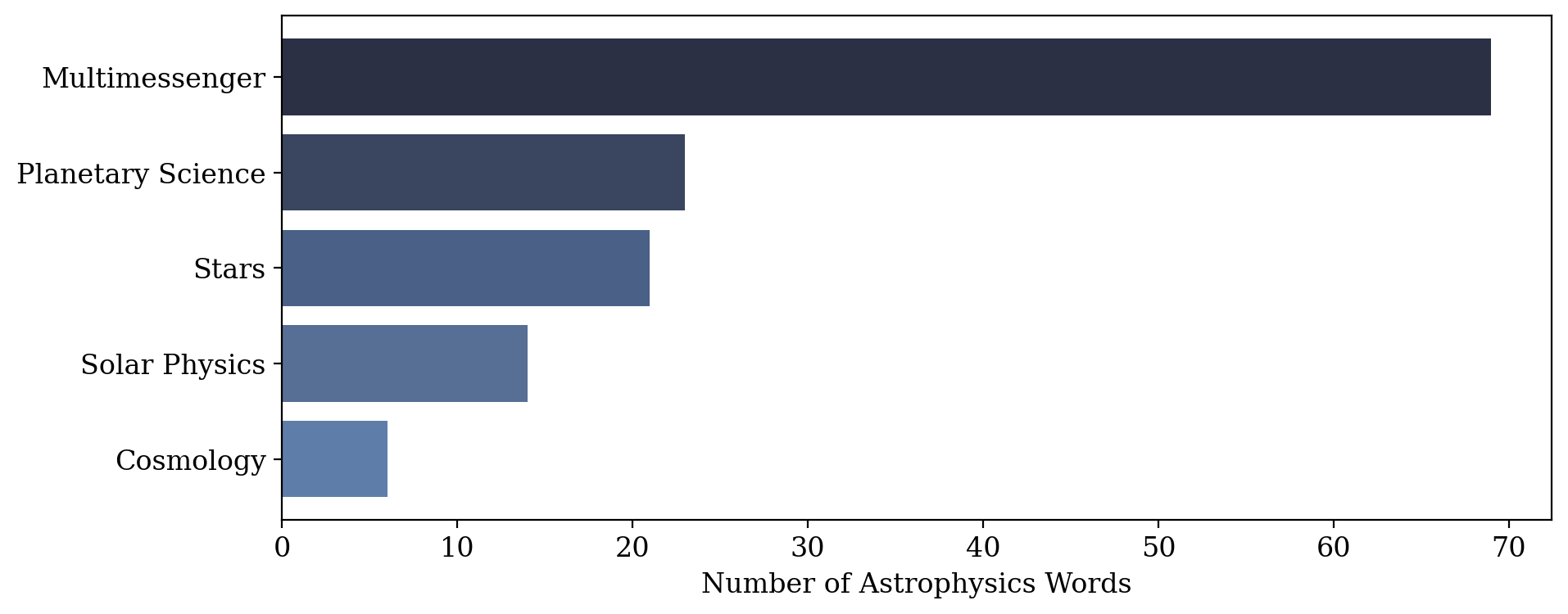}
    \caption{Astrophysics fields that had the most words throughout Swift's discography \textit{and} played during the Eras tour.}
    \label{fig:fields}
\end{figure*}

We then seek to find which field of astrophysics Taylor Swift promoted the most throughout the Eras tour. To do this, we iterate through each song's astrophysics words, checking which category in Figure \ref{fig:words} they belong to. Then we count the occurences of words in each category in a similar way as in Figure \ref{fig:tour_score} and plot the results in Figure \ref{fig:fields}. We find Swift to be most interested in the field of multimessenger astronomy, with an APS close to 2000. Other topics in astronomy (namely planetary science, stars and solar physics) show a similar APS between 250-500, with cosmology being by far the field with the lowest APS score. This could be due to personal interest, or because her latest albums were released before the latest cosmology results \citep{desi, ACT, KiDS}. In case it is due to the latter, we might see a shift in this result in her next album.

We also found that the fields promoted the most did not change when considering Swift's entire discography, rather than just those played the most at the Eras tour, proving this focused analysis is a good representation of her complete discography.

\section{A Taylor (Swift) expansion}
\label{sec:taylorexpansion}
From our data, we explored whether we can determine the number of unique astrophysics words mentioned by an album in Swift's discography as a function of the album number. We aim to compute this by carrying out a Taylor series expansion. For a function, this is defined as an infinite sum of terms expressed in terms of the function's derivatives around a point.

\begin{equation}
f(x) = \sum_{n=0}^{\infty} \frac{f^{(n)}(a)}{n!} (x - a)^n
\end{equation}

\begin{figure}
    \centering
    \includegraphics[width=0.5\textwidth]{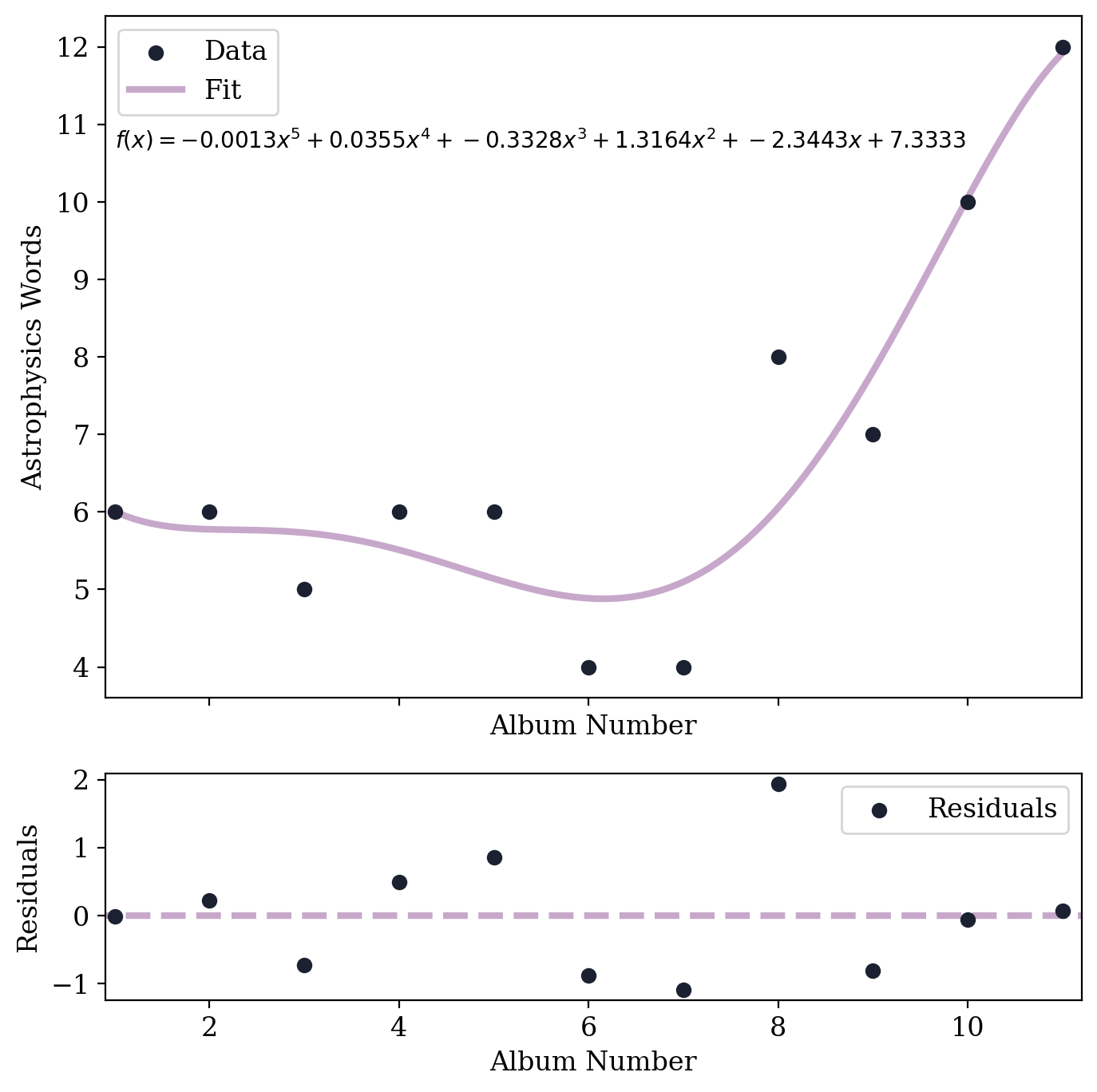}
    \caption{Taylor expansion of the number of unique astrophysics songs as a function of album number}
    \label{fig:expansion}
\end{figure}

In order to calculate this, we plot the number of physics words as a function of the album number, and first attempt a simple polynomial fit. We see that our data can be approximated by a quintic polynomial, which also serves as our expansion. This is shown in Figure \ref{fig:expansion}
We can then use this polynomial to predict the number of astrophysics words in her upcoming album. As we highlighted in Section \ref{sec:results1}, there is an increasing trend in the number of unique words per album in her most recent releases. According to our Taylor expansion, we expect $12 \pm 5$ unique astrophysics words in her next release, making it the most astrophysical album to date. Future work could use future upcoming releases to improve this approximation, as well as more complex fits to the data than a polynomial function (i.e. by using symbolic regression).

\section{Look What You Made Me Do}
\label{sec:kanye}

Given that Swift has proven to be a strong advocate for astrophysics, particularly in the realm of multimessenger astronomy, it raises the question of whether other artists similarly encourage their young audiences to pursue careers in science. For no reason at all \citep{west1,west2}, we decided to explore the discography of Kanye West.

Since \textit{The Tortured Poets Department}, Swift’s latest album, contained the highest number of unique astrophysical terms (as highlighted in Section \ref{sec:results1}), we chose to compare it with West’s most recent solo album, Donda (released in 2021). With 27 tracks, Donda is similar in length to \textit{The Tortured Poets Department}, which has 31 songs. We identified 6 astrophysics-related terms in Donda: ‘light,’ ‘sparkle,’ ‘moon,’ ‘interstellar,’ ‘infrared,’ and ‘stars’, clearly showing that other artists have interests in similar areas of astrophysics as Swift (stellar astrophysics, multimessenger astronomy and planetary sciences), this is half the number of words found in \textit{The Tortured Poets Department}. Given that both albums have a similar length, from this we conclude that West does not place the same emphasis on promoting astrophysics research as Swift. Moreover, the language used in Donda is not age appropriate or conducive to inspiring young listeners to pursue science.

\section{It's me hi, I'm the problem, it's me}
\label{sec:discussion}

Whilst it is clear from Figures \ref{fig:albums_vs_words} and \ref{fig:tour_score} that Swift promotes the study of astrophysics both in her total discography and throughout the Eras tour, there are some fields of astrophysics research that cannot be found within her lyrics. 

For example, gravitational lensing is not a concept that features in Swift's work. Strong lensing is a powerful astrophysical tool that helps us detect dark, massive substructures \citep{dan}, constrain the Hubble tension with the time delays of lensed supernovae \citep{ana}, understand dark energy \citep{tian}, and measure stellar mass \citep{luke}, making it a valuable technique in modern cosmology. 

Another field of study that has been neglected, much like the album \textit{evermore}, are the strong emission lines from young stellar populations observed at high redshift by JWST \citep{cameron,boyett} which are crucial to include when forward modelling cosmological simulations or performing SED fitting \citep[`inverse modelling',][]{sophie}.

Moreover, as seen in Figure \ref{fig:fields} and discussed in Section \ref{sec:eras_tour_promotion}, cosmology is not the main focus of her music yet, but her upcoming album could be motivated by the most recent cosmological results, which include values for the parametrisation of the equation of state of dark energy and a reanalysis of the $S_8$ tension. 

We therefore think it a good idea to suggest lyrics to Swift to incorporate these new fields of study into her next work and to expand on current topics, and these suggestions are shown below for each field:

\subsection*{Strong lensing}

\begin{enumerate}
    \item Through your gravity, I curve, a path I can't control
    \item Lensed light, rewriting fate. Every delay, a clue to space
    \item Lensed through your eyes, I see what’s concealed
    \item You let me see what no one can trace, The hidden masses, the secrets in space.
    \item Through the lens, you shape the way, Dark substructures in the Milky Way.
    \item And if you could bend, just one more time, Could we fix the tension in the skies?
\end{enumerate}

\subsection*{High redshift emission lines}

\begin{enumerate}
    \item H$\beta$ calling, forbidden lines trace, Through dust and gas in a faraway place.
    \item Oxygen sings, hydrogen cries, A spectral echo left in the sky.
    \item Falling light at redshift z, Telling stories left for me.
    \item Inverse modeling can’t find the way, If the code is wrong, if the past won’t obey.
    \item We build a model, we fit the light, Forward in time, but nothing’s right
    \item You left long ago, but your glow remains, Whispering secrets in Lyman’s name.
\end{enumerate}

\subsection*{Cosmology}
\begin{enumerate}
    \item We swore we’d stay, $\Omega$ close to one,
But acceleration’s won, and we’re undone
    \item Dark energy’s fading, I feel you near,
The universe slows, but your voice stays clear.
    \item Thought we were doomed to drift apart,
But the vacuum’s losing its hold on my heart.
    \item We were chasing shadows in the dark,
But $\Lambda$CDM still holds its spark.
\end{enumerate}
\section{Conclusions (10 minute version)}
\label{sec:conclusions}

This study highlights the surprising and impactful intersection between Taylor Swift's discography and astrophysics, demonstrating how her lyrics promote public engagement in the fields of solar physics, cosmology, stellar astrophysics, planetary science, and multimessenger astronomy. Our specific conclusions are as follows:

\begin{itemize}
    \item Swift's most recent album, \textit{The Tortured Poets Department} promotes astrophysics the most, with \textit{Lover} and \textit{Reputation} promoting the least. This change in Swift's interest in the field is likely due to recent developments and advancements in the field of astrophysics and cosmology, as well as the increased productivity of researchers.
    
    \item Throughout Swift's Eras tour, \textit{Midnights} promotes astrophysics to its audience the most partially due to the \textit{Midnights} set list being longer than those of Swift's other albums.

    \item Taylor Swift is a big fan of multimessenger astronomy and is excited to share to her fans how combining information from different signals is an exciting prospect. 

    \item We perform a Taylor (Swift) expansion and predict $12 \pm 5$ unique astrophysics words in her next release, making it the most astrophysics intense album to date.

    \item We compare the number of unique astrophysics terms in Swift's last album and that by Kanye West and find that Swift promotes astrophysics to her fans to a higher degree.

    \item We find that fields such as strong gravitational lensing and the study of high redshift emission lines are not common themes in Swift's work and suggest relevant lyrics for her next work.
\end{itemize}

Taylor Swift’s exceptional contributions to astrophysics communication are poised to have a lasting, positive impact on the field, inspiring the next generation of astrophysicists through her lyrics. In fact, the number of astronomy majors in the US has been steadily increasing over the last few decades, rising over 300\% from 20 years ago. Coincidentally, her first album was released in 2006. The area of astronomy undoubtedly has Swift to thank for this increase in students.

While her Eras Tour shattered records with unprecedented attendance, boosted craft sales, and became the highest-grossing concert film of all time, her influence extends beyond music. Swift’s ability to engage and educate through storytelling sets a remarkable example in science communication, making astrophysics more accessible and captivating to a broader audience.

\textbf{We are never, ever, ever getting published together.}

\section*{Acknowledgements}

The authors extend their gratitude to Nathan Cruickshank for giving us the idea to use a Taylor Swift lyric searcher and to Jacob Newman for generously providing many of the Taylor Swift puns featured in this article.

The authors give a huge congratulations to their friends and fellow PhD students Susanna Green and Joseph Callow on the recent submission of their PhD theses.

%%%%%%%%%%%%%%%%%%%%%%%%%%%%%%%%%%%%%%%%%%%%%%%%%%
\section*{Data Availability}

We have made the data compiled in this study available \href{https://docs.google.com/spreadsheets/d/1lecR6zBvhs-hZFn5VoqDMUTvjdoznoXkxlZ8PSrIqN8/edit?usp=sharing}{here}, allowing readers to explore which astrophysical terms appear in each song. If readers have suggestions for additional terms or data, they are encouraged to contact the authors.

%%%%%%%%%%%%%%%%%%%% REFERENCES %%%%%%%%%%%%%%%%%%

% The best way to enter references is to use BibTeX:

%\bibliographystyle{mnras}
%\bibliography{bib} % if your bibtex file is called example.bib

% Alternatively you could enter them by hand, like this:
% This method is tedious and prone to error if you have lots of references

%%%%%%%%%%%%%%%%%%%%%%%%%%%%%%%%%%%%%%%%%%%%%%%%%%

%%%%%%%%%%%%%%%%% APPENDICES %%%%%%%%%%%%%%%%%%%%%

\appendix

\section{(From the vault)}

\begin{figure*}
    \centering
    \includegraphics[width=\textwidth]{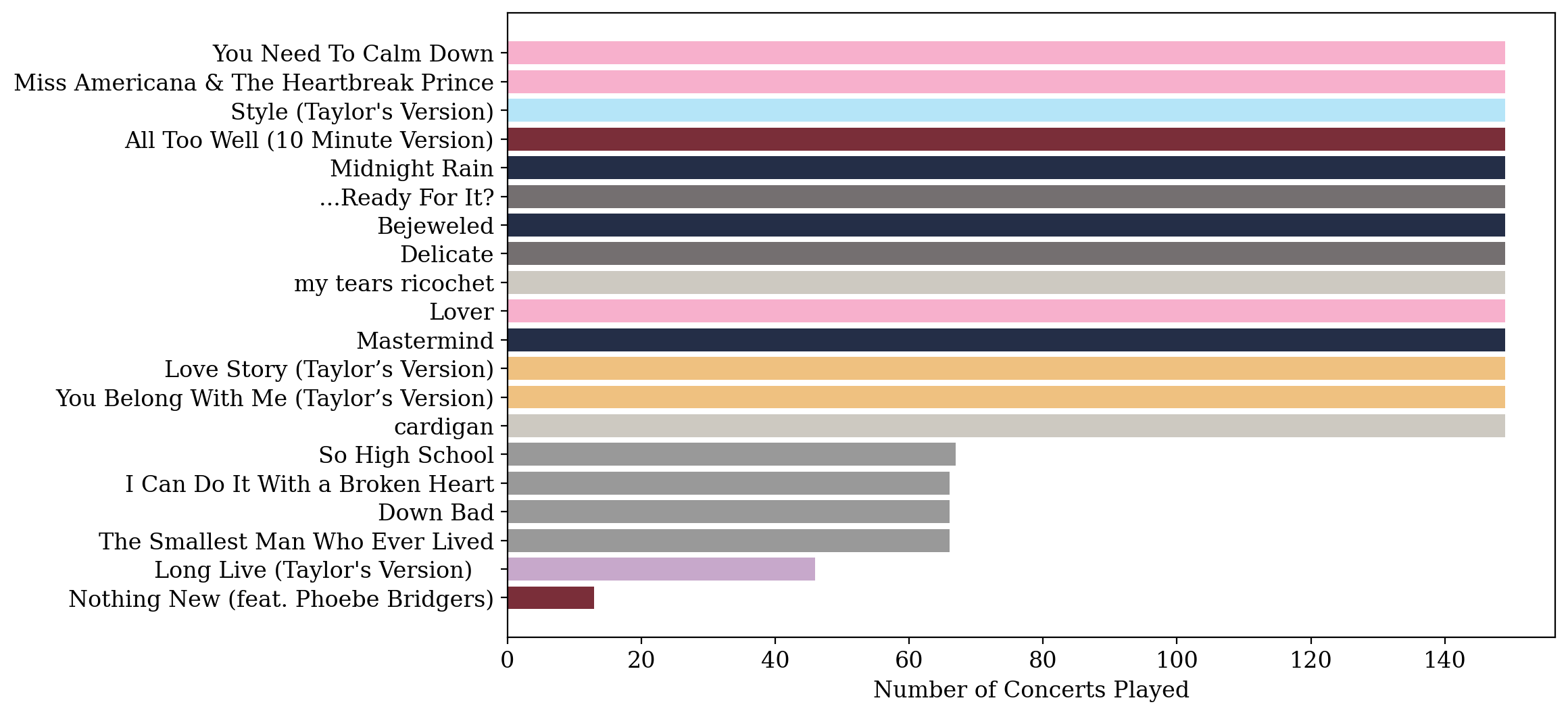}
    \caption{Swift's twenty most played songs at the Eras tour which contain astrophysics terms.}
    \label{fig:top_20}
\end{figure*}

For interested readers, Figure \ref{fig:top_20} displays the top 20 songs in Swift's discography that include astrophysical terms, ranked by their frequency of play during the Eras Tour. Notably, the top 14 songs were performed at every concert, reaching the maximum possible count of 149 performances.

%%%%%%%%%%%%%%%%%%%%%%%%%%%%%%%%%%%%%%%%%%%%%%%%%%

% Don't change these lines
\bsp	% typesetting comment
\label{lastpage}
\end{document}